\documentclass[twocolumn,showpacs,showkeys,amsmath,amssymb]{revtex4}

\usepackage{graphicx}

\begin{document}

\title{Gravitational lensing in the strong field limit}

\author{V. Bozza}

\email{valboz@sa.infn.it}

\affiliation{Dipartimento di Fisica ``E.R. Caianiello'',
Universit\`a di Salerno, Italy.\\
   Istituto Nazionale di Fisica Nucleare, Sezione di
 Napoli.}

\date{\today}

\begin{abstract}
We provide an analytic method to discriminate among different
types of black holes on the ground of their strong field
gravitational lensing properties. We expand the deflection angle
of the photon in the neighbourhood of complete capture, defining a
strong field limit, in opposition to the standard weak field
limit. This expansion is worked out for a completely generic
spherically symmetric spacetime, without any reference to the
field equations and just assuming that the light ray follows the
geodesics equation. We prove that the deflection angle always
diverges logarithmically when the minimum impact parameter is
reached. We apply this general formalism to Schwarzschild,
Reissner-Nordstrom and Janis-Newman-Winicour black holes. We then
compare the coefficients characterizing these metrics and find
that different collapsed objects are characterized by different
strong field limits. The strong field limit coefficients are
directly connected to the observables, such as the position and
the magnification of the relativistic images. As a concrete
example, we consider the black hole at the centre of our galaxy
and estimate the optical resolution needed to investigate its
strong field behaviour through its relativistic images.
\end{abstract}



\maketitle

\section{Introduction}

Gravitational lensing is one of the first applications of General
Relativity ever studied \cite{Einstein}. Firstly it was recognized
in the light deflection by sun, secondly in lensing of quasars by
foreground galaxies, then in the formation of giant arcs in galaxy
clusters and finally in galactic microlensing. Now it is an
ordinary phenomenon in the panorama of astronomical observations
(see \cite{SEF} for a complete treatment and references therein).

The full theory of gravitational lensing has been developed
following the scheme of the weak field approximation and, in this
formulation, it has been successfully employed to explain all the
physical observations.

In the last years, however, the scientific community is starting
to look at this phenomenon from the opposite point of view,
opening a strong field perspective. Viergutz \cite{Vie} made a
semi-analytical investigation about geodesics in Kerr geometry; in
Ref. \cite{Bar} the appearance of a black hole in front of a
uniform background was studied; Falcke, Melia and Agol \cite{FMA}
considered the emission of the accretion flow as source. Virbhadra
\& Ellis \cite{VirEll} showed that a source behind a Schwarzschild
black hole would produce one set of infinite relativistic images
on each side of the black hole. These images are produced when a
light ray with small impact parameter winds one or several times
around the black hole before emerging. Later on, by an alternative
formulation of the problem, Frittelli, Kling \& Newman \cite{FKN}
attained an exact lens equation, giving integral expressions for
its solutions, and compared their results to those by Virbhadra \&
Ellis. The same problem has been investigated by Bozza et al. in
Ref. \cite{BCIS}, where a strong field limit was first defined in
Schwarzschild black hole lensing and used to find the position and
the characteristics of all the relativistic images analytically.
Eiroa, Romero and Torres \cite{ERT} applied the same technique to
a Reissner - Nordstrom black hole. Recently, in another work
\cite{VirEll2}, Virbhadra \& Ellis distinguished the main features
of gravitational lensing by normal black holes and by naked
singularities, analyzing the Janis, Newman, Winicour metric. They
remarked the importance of these studies in providing a test for
the cosmic censorship hypothesis.

The reason for such an interest in gravitational lensing in strong
fields is that by the properties of the relativistic images it may
be possible to investigate the regions immediately outside of the
event horizon. High resolution imaging of black holes by VLBI
\cite{VLBI} could be able to detect the relativistic images and
retrieve information about strong fields stored within these new
observables. Moreover, since alternative theories of gravitation
must agree with GR in the weak field limit, in order to show
deviations from GR it is necessary to probe strong fields in some
way. Indeed, deviation of light rays in strong fields is one of
the most promising grounds where a theory of gravitation can be
tested in its full form.

Of course, the study of null geodesics in strong fields is not
easy and up to now it has always been carried out using numerical
techniques. An analytical treatment would enlighten the dependence
of the observables on the parameters of the system, allow easy
checks about the detectability of the images and open the way to
comparisons between the results in different metrics. In Ref.
\cite{BCIS}, a new way to expand the deflection angle in the
Schwarzschild metric was suggested. The deflection angle near its
divergence was approximated by its leading order and its first
regular term and then plugged into the lens equation. In this way,
very simple and reliable analytical formulae were derived for the
relativistic images and their main features.

So, as the weak field limit takes the first order deviation from
Minkowski, the strong field limit starts from complete capture of
the photon and takes the leading order in the divergence of the
deflection angle.

The strong field limit of Ref. \cite{BCIS} was only developed in
Schwarzschild spacetime. In this paper, we provide a general
method to extend the strong field limit to a generic static
spherically symmetric spacetime. Our method is universal and can
be applied to any spacetime in any theory of gravitation, provided
that photons satisfy the standard geodesics equation. The
parameters of the strong field limit expansion are directly
connected with the observables, providing an effective tool to
discriminate among different metrics. In Sect. 2, we state the
problem and carry out the strong field limit of the deflection
angle. In Sect. 3, we apply the method to some simple metrics:
Schwarzschild, Reissner-Nordstrom and Janis-Newman-Winicour black
hole, discussing their differences with reference to the
gravitational lensing phenomenology. In Sect. 4, we establish a
connection between the strong field limit coefficients and the
relativistic images, analyzing the case of the black hole at the
center of our galaxy as a concrete example where our results can
be tested. Finally, Sect. 5 contains the summary.

\section{Strong field expansion of the deflection angle}
\label{Sec Strong Field}

A generic spherically symmetric spacetime has line-element
\cite{Wei}
\begin{equation}
ds^2=A(x)dt^2-B(x)dx^2-C(x)\left( d\theta^2+\sin^2\theta \phi^2
\right).
\end{equation}
where
\begin{equation}
\begin{array}{c}
  A(x) \longrightarrow^{\hspace{-.7cm} x\rightarrow \infty} 1 - \frac{2M}{x}
  \\ \\
  B(x) \longrightarrow^{\hspace{-.7cm} x\rightarrow \infty} 1 +
  \frac{2M}{x} \\ \\
  C(x) \longrightarrow^{\hspace{-.7cm} x\rightarrow \infty} x^2,
\end{array}
\end{equation}
in order to correctly match the weak gravitational field far from
the lensing object.

We require that the equation
\begin{equation}
\frac{C'(x)}{C(x)}=\frac{ A'(x)}{A(x)} \label{Eqm}.
\end{equation}
admits at least one positive solution. We shall call the largest
root of Eq. (\ref{Eqm}) the radius of the photon sphere $x_m$ (for
an alternative definition of the photon sphere, see \cite{CVE}).
$A$, $B$, $C$, $A'$ and $C'$ must be positive for $x>x_m$

For metrics expressed in standard coordinates ($C(x)=x^2$) a
sufficient condition for the existence of $x_m$ is the presence of
a static limit (a radius $x_s$ such that $A(x_s)=0$). Our strong
field expansion takes the photon sphere as the starting point. In
our study, therefore, we shall not consider naked singularities
without a photon sphere. For a numerical study of this situation,
see Ref. \cite{VirEll2}.

A photon incoming from infinity with some impact parameter $u$,
will be deviated while approaching the black hole. It will reach a
minimum distance $x_0$ and then emerge in another direction. Of
course, the approach phase is identical to the departure phase,
with the time reversed.

By the conservation of the angular momentum, the closest approach
distance is related to the impact parameter $u$ by
\begin{equation}
u=\sqrt{\frac{C_0}{A_0}}. \label{Imppar-mindis}
\end{equation}
where the subscript $0$ indicates that the function is evaluated
at $x_0$.

From the geodesics equation it is easy to extract the quantity
\begin{equation}
\frac{d\phi}{dx}=\frac{\sqrt{B}}{\sqrt{C}
\sqrt{\frac{C}{C_0}\frac{A_0}{A}-1}}
\end{equation}
which gives the angular shift of the photon as a function of the
distance from the center (see \cite{Wei} for the complete
derivation).

The deflection angle can then be calculated as a function of the
closest approach:
\begin{eqnarray}
&& \alpha(x_0)=I(x_0)-\pi \\ %
&& I(x_0)=\int\limits_{x_0}^\infty \frac{2\sqrt{B}dx}{\sqrt{C}
\sqrt{\frac{C}{C_0}\frac{A_0}{A}-1}} \label{I x}.
\end{eqnarray}
It is easy to check that for a vanishing gravitational field
($A=B=1$, $C=x^2$) $\alpha(x_0)$ identically vanishes. In the weak
field limit, the integrand is expanded to the first order in the
gravitational potential. This limit is no longer valid when the
closest approach distance significantly differs from the impact
parameter (which, by Eq. (\ref{Imppar-mindis}), means that
$A(x_0)$ significantly differs from 1 and/or $C(x_0)$ from
$x_0^2$, i.e. the photon passes in a strong gravitational field).

\begin{figure}
\resizebox{\hsize}{!}{\includegraphics{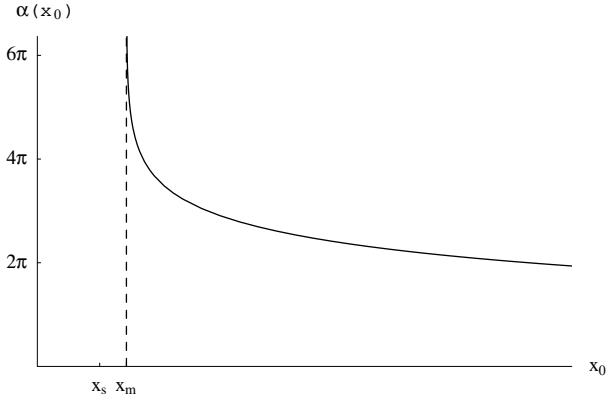}}
 \caption{General behaviour of the deflection angle %
 as a function of the closest approach $x_0$. The %
 deflection angle increases as $x_0$ decreases and diverges %
 at $x_0=x_m$. Each time $\alpha(x_0)$ reaches a multiple of %
 $2\pi$, the photon completes a loop before emerging.}
 \label{Fig deflection angle}
\end{figure}

When we decrease the impact parameter (and consequently $x_0$),
the deflection angle increases. At some point, the deflection
angle will exceed $2\pi$, resulting in a complete loop around the
black hole. Decreasing $u$ further, the photon will wind several
times before emerging. Finally, for $x_0=x_m$ (see Sect. \ref{Sect
a}), corresponding to an impact parameter $u=u_m$, the deflection
angle diverges and the photon is captured (see Fig. \ref{Fig
deflection angle}).

In this section, we will show that this divergence is logarithmic
for all spherically symmetric metrics. Our purpose is to get an
analytical expansion of the deflection angle close to the
divergence in the form
\begin{equation}
\alpha(x_0)=-a \log \left(\frac{x_0}{x_m}-1 \right) +b
+O\left(x_0-x_m \right), \label{S F L}
\end{equation}
where all the coefficients depend on the metric functions
evaluated at $x_m$.

As the angular separation of the image from the lens is
$\theta=\frac{u}{D_{OL}}$, where $D_{OL}$ is the distance between
the lens and the observer, we need to express the deflection angle
in terms of this variable. So, we shall finally transform
Eq.(\ref{S F L}) to
\begin{equation}
\alpha(\theta)=-\overline{a} \log \left(\frac{\theta
D_{OL}}{u_m}-1 \right) +\overline{b} +O\left(u-u_m \right)
\label{S F L theta}
\end{equation}
which we define as the {\it strong field limit} of the deflection
angle. The rest of this section is devoted to the calculation of
the two coefficients $a$ and $b$ in a first step and then of
$\overline{a}$ and $\overline{b}$.

\subsection{Divergent term of the deflection angle} \label{Sect a}

We define two new variables
\begin{eqnarray}
&& y=A(x) \\%
&& z= \frac{y-y_0}{1-y_0}
\end{eqnarray}
where $y_0=A_0$. The integral (\ref{I x}) in the deflection angle
becomes
\begin{eqnarray}
&& I(x_0)=\int\limits_0^1 R(z,x_0) f(z,x_0) dz \label{I z} \\%
&& R(z,x_0)=\frac{2\sqrt{B y}}{C A' }\left( 1-y_0 \right)
\sqrt{C_0} \label{R} \\%
&& f(z,x_0)=\frac{1}{\sqrt{y_0- \left[ \left(1-y_0 \right) z+ y_0
\right]\frac{C_0}{C}}}
\end{eqnarray}
where all functions without the subscript $0$ are evaluated at
$x=A^{-1} \left[\left(1-y_0 \right) z+ y_0 \right]$.

The function $R(z,x_0)$ is regular for all values of $z$ and
$x_0$, while $f(z,x_0)$ diverges for $z \rightarrow 0$. To find
out the order of divergence of the integrand, we expand the
argument of the square root in $f(z,x_0)$ to the second order in
$z$:
\begin{eqnarray}
& f & (z,x_0) \sim f_0(z,x_0)= \frac{1}{\sqrt{\alpha z +\beta
z^2}}
\\%
& \alpha & =\frac{1-y_0}{C_0 A'_0}  \left(C'_0 y_0-C_0 A'_0
\right) \label{alpha} \\%
& \beta & = \frac{\left( 1-y_0 \right)^2}{2C_0^2 {A'_0}^3} \left[
2C_0 C'_0 {A'_0}^2 + \right.  \nonumber
\\%
&&\left. +\left(C_0 C''_0-2{C'_0}^2 \right)y_0 A'_0 -C_0 C'_0 y_0
A''_0 \right]. \label{beta}
\end{eqnarray}

When $\alpha$ is non zero, the leading order of the divergence in
$f_0$ is $z^{-1/2}$, which can be integrated to give a finite
result. When $\alpha$ vanishes, the divergence is $z^{-1}$ which
makes the integral diverge. Examining the form of $\alpha$, we see
that it vanishes at $x_0=x_m$, with $x_m$ defined by Eq.
(\ref{Eqm}). Each photon having $x_0<x_m$ is captured by the
central object and cannot emerge back.

To solve the integral (\ref{I z}), we split it into two pieces
\begin{equation}
I(x_0)=I_D(x_0)+I_R(x_0),
\end{equation}
where
\begin{equation}
I_D(x_0) = \int\limits_0^1 R(0,x_m) f_0(z,x_0) dz
\end{equation}
contains the divergence and
\begin{equation}
I_R(x_0) = \int\limits_0^1 g(z,x_0)dz \label{IR def}
\end{equation}
\begin{equation}
g(z,x_0)= R(z,x_0)f(z,x_0)-R(0,x_m)f_0(z,x_0).
\end{equation}
is the original integral with the divergence subtracted. We shall
solve both integrals separately and then sum up their results to
rebuild the deflection angle. Here we deal with $I_D$ and its
divergence, while in the next subsection we shall verify that
$I_R$ is indeed regular.

The integral $I_D(x_0)$ can be solved exactly, giving
\begin{equation}
I_D(x_0)=R(0,x_m)\frac{2}{\sqrt{\beta}}\log
\frac{\sqrt{\beta}+\sqrt{ \alpha+\beta} }{ \sqrt{\alpha} }.
\end{equation}

Since we are interested in the terms up to $O(x_0-x_m)$, we expand
$\alpha$ as
\begin{equation}
\alpha=\frac{2 \beta_m A'_m}{1-y_m} \left( x_0-x_m \right) +
O\left( x_0-x_m \right)^2,
\end{equation}
where
\begin{equation}
\beta_m=\beta|_{x_0=x_m}=\frac{ C_m \left( 1- y_m \right)^2
\left(C''_m y_m-C_m A''(x_m) \right)}{2y_m^2 {C'_m}^2}
\label{betam}
\end{equation}
and substitute in $I_D(x_0)$. Rearranging all terms, we find
\begin{equation}
I_D(x_0)=-a \log
\left(\frac{x_0}{x_m}-1 \right)+b_D + O(x_0-x_m), \label{ID}
\end{equation}
\begin{eqnarray}
&& a=\frac{R(0,x_m)}{\sqrt{\beta_m}} \label{a} \\ %
&& b_D=\frac{R(0,x_m)}{\sqrt{\beta_m}} \log \frac{2(1-y_m)}{A'_m
x_m} \label{bD}.
\end{eqnarray}
Eq. (\ref{ID}) yields the leading order in the divergence of the
deflection angle, which is logarithmic, as anticipated before. The
coefficient $a$ of Eq. (\ref{S F L}) is then given by Eq.
(\ref{a}).

\subsection{Regular term of the deflection angle} \label{Sect b}

In order to find the correct coefficient $b$ in Eq. (\ref{S F L}),
we have to add to the term $b_D$ coming from Eq. (\ref{bD}), an
analogous term coming from the regular part of the original
integral, defined by Eq. (\ref{IR def}).

We can expand $I_R(x_0)$ in powers of $(x_0-x_m)$
\begin{equation}
I_R(x_0)=\sum\limits_{n=0}^\infty \frac{1}{n!}
(x_0-x_m)^n\int\limits_0^1 \left. \frac{\partial^n g}{\partial
x_0^n} \right|_{x_0=x_m} dz
\end{equation}
and evaluate the single coefficients.

If we had not subtracted the singular part from $R(z,x_0)
f(z,x_0)$, we would have an infinite coefficient for $n=0$, while
all other coefficients would be finite. However, the function
$g(z,x_0)$ is regular in $z=0,x_0=x_m$ as can be easily checked by
a power expansion, recalling that $\alpha_m=0$.

Since we are interested to terms up to $O(x_0-x_m)$, we will just
retain the $n=0$ term
\begin{equation}
I_R(x_0)=\int\limits_0^1 g(z,x_m) dz +O(x_0-x_m) \label{IR}
\end{equation}
and then
\begin{equation}
b_R=I_R(x_m) \label{bR}
\end{equation}
is the term we need to add to $b_D$ in order to get the regular
coefficient. Recalling also the term $-\pi$ in the deflection
angle, we have
\begin{equation}
b=-\pi+b_D+b_R.
\end{equation}

The coefficient $b_R$ can be easily evaluated numerically for all
metrics, since the integrand has no divergence. However, in many
cases it is also possible to build a completely analytical formula
for $b_R$ as well. In fact, in Schwarzschild metric, the integral
(\ref{IR}) is solved exactly (see Sect. \ref{Sec sch}). Then, in
most metrics, we can expand Eq. (\ref{IR}) in powers of their
parameters, starting from the Schwarzschild limit, and evaluate
each coefficient separately. This is what we shall do for
Reissner-Nordstrom and Janis-Newman-Winicour black holes (see
Sect. \ref{Sec Applications}).

\subsection{From $\alpha(x_0)$ to $\alpha(\theta)$}

From Eq. (\ref{Imppar-mindis}), we see that the minimum impact
parameter is
\begin{equation}
u_m=\sqrt{\frac{C_m}{y_m}}. \label{um}
\end{equation}
Expanding Eq. (\ref{Imppar-mindis}), we find
\begin{eqnarray}
&& u-u_m=c \left(x_0-x_m \right)^2 \\%
&& c=\frac{C''_m y_m - C_m A''_m}{4 \sqrt{y_m^3 C_m}}=\beta_m
\sqrt{\frac{y_m}{C_m^3}} \frac{{C'_m}^2}{2\left(1-y_m \right)^2}.
\end{eqnarray}

Using this relation, we can write the deflection angle as a
function of $\theta$:

\begin{eqnarray}
&& \alpha(\theta)=-\overline{a} \log \left( \frac{\theta
D_{OL}}{u_m} -1
\right) +\overline{b} \label{alpha theta}\\%
&& \overline{a}=\frac{a}{2}= \frac{R(0,x_m)}{2\sqrt{\beta_m}} \label{oa}\\%
&& \overline{b}=b+\frac{a}{2} \log{\frac{c x_m^2}{
u_m}}=-\pi+b_R+\overline{a}\log{\frac{2\beta_m}{y_m}}.\label{ob}
\end{eqnarray}

This concludes our general discussion of the form of the
deflection angle in the strong field limit. Even if the proof is
somewhat tricky, the application to concrete cases is very
straightforward, as we shall see in Sect. \ref{Sec Applications}.
In fact, once we write a metric, it is sufficient to:

\begin{itemize}
\item solve Eq. (\ref{Eqm}) to find $x_m$;
\item write $\beta_m$ from Eq. (\ref{betam}) and $R(0,x_m)$ from
Eq. (\ref{R});
\item compute $b_R$ from Eq. (\ref{bR}) numerically or by a proper expansion in the parameters
of the metric;
\item compute the coefficients $u_m$, $\overline{a}$ and $\overline{b}$
from Eqs. (\ref{um}), (\ref{oa}) and (\ref{ob}), respectively.
\end{itemize}

The crucial step is the calculation of $b_R$, since it is the only
integral involved in the whole procedure.

\section{Applications} \label{Sec Applications}

In this section, we apply the general formulation of the strong
field limit to three simple examples. First, we shall revisit
Schwarzschild spacetime, whose strong field limit gravitational
lensing has already been studied in Ref. \cite{BCIS}. Then we will
apply our method to Reissner-Nordstrom spacetime, which was
explored numerically in Ref. \cite{ERT}. Finally, as an example of
extended theory of gravitation, we shall consider a Janis, Newman,
Winicour black hole, whose strong field limit expansion has not
been investigated so far. In all three cases, we shall derive
analytical formulae for the strong field limit coefficients, in
order to analyze the functional dependences of the deflection
angle on the parameters of the metrics.

\subsection{Schwarzschild lensing} \label{Sec sch}

This is the simplest spherically symmetric metric describing the
outer solution for a black hole. It only depends on the mass of
the central object (by Birkhoff's theorem). It is convenient to
define the Schwarzschild radius $x_s=2M$ as the measure of
distances; then, in standard coordinates, the functions in the
metric become
\begin{eqnarray}
&&A(x)= 1- \frac{1}{x} \\ %
&&B(x)= \left( 1- \frac{1}{x} \right)^{-1} \\%
&&C(x)=x^2,
\end{eqnarray}
which obviously satisfy all hypotheses required in Sect. \ref{Sec
Strong Field}, with static limit $x_s=1$.

The two functions $R(z,x_0)$ and $f(z,x_0)$ read
\begin{eqnarray}
&&R(z,x_0)=2 \\%
&&f(z,x_0)=\frac{1}{\sqrt{\left( 2-\frac{3}{x_0} \right) z+ \left(
\frac{3}{x_0} -1 \right) z^2-\frac{z^3}{x_0}}}.
\end{eqnarray}

From Eqs. (\ref{alpha}), (\ref{beta}), or directly from the
expansion of the denominator of $f$, we read off the coefficients
$\alpha$ and $\beta$
\begin{eqnarray}
&& \alpha= 2-\frac{3}{x_0} \label{alpha sch}\\%
&& \beta= \frac{3}{x_0} -1. \label{beta sch}
\end{eqnarray}

The equation $\alpha=0$ defines the radius of the photon sphere
\begin{equation}
x_m=\frac{3}{2}.
\end{equation}
Consequently, $\beta_m=1$.

In this simple case, it is possible to solve the integral in Eq.
(\ref{bR}) exactly and write the regular term in the deflection
angle
\begin{equation}
b_R=2 \log \left[ 6 \left( 2-\sqrt{3} \right) \right]=0.9496.
\label{bR Sch}
\end{equation}

From Eqs. (\ref{oa}) and (\ref{ob}), we derive the coefficients
$\overline{a}$, $\overline{b}$ and $u_m$ of the deflection angle
\begin{eqnarray}
&& \overline{a}=1 \\ %
&& \overline{b}=-\pi+b_R+\log 6=-0.4002 \\%
&& u_m=\frac{3 \sqrt{3}}{2}.
\end{eqnarray}

Then the Schwarzschild deflection angle, in the strong field
limit, is
\begin{equation}
\alpha(\theta)=-\log \left(\frac{2\theta D_{OL}}{3\sqrt{3}} -1
\right) + \log \left[ 216 \left( 7-4\sqrt{3} \right) \right]-\pi,
\end{equation}
as obtained in Ref. \cite{BCIS}.

In this simple case, we can compare the exact deflection angle
$\alpha_{ex}(\theta)$, calculated numerically and the strong field
limit $\alpha_{app}(\theta)$. The most external image appears
where $\alpha(\theta)$ falls below $2\pi$. This happens at
$u-u_m=0.003264$. Here the discrepancy between $\alpha_{ex}$ and
$\alpha_{app}$ is about $0.06\%$, which corresponds to an error in
the position of the outer image of the order $0.4\%$. With such a
high accuracy, we are encouraged to take the Schwarzschild strong
field limit as the starting point for successive series expansions
to evaluate $b_R$ in more advanced metrics.

\subsection{Reissner-Nordstrom lensing}

Reissner-Nordstrom metric describes the gravitational field of a
spherically symmetric massive object endowed with an electric
charge $q$. The metric functions in standard coordinates are
\begin{eqnarray}
&&A(x)= 1- \frac{1}{x}+\frac{q^2}{x^2} \\ %
&&B(x)= \left( 1- \frac{1}{x}+\frac{q^2}{x^2} \right)^{-1} \\%
&&C(x)= x^2 .
\end{eqnarray}
They satisfy the hypotheses required in Sect. \ref{Sec Strong
Field}, only when $q \leq \frac{3}{4\sqrt{2}}$. However, beyond
the critical value $q=0.5$, there is no event horizon and
causality violations appear \cite{Carter,HawEll}. We shall
restrict to $q<0.5$.

The coefficients $\alpha$ and $\beta$ are
\begin{eqnarray}
&& \alpha=\left( 2-\frac{3}{x_0} +\frac{4q^2}{x_0^2} \right)
\frac{x_0-q^2}{x_0-2q^2} \\%
&& \beta= \left( \frac{3}{x_0} -1 -\frac{9q^2}{x_0^2}
+\frac{8q^4}{x_0^3} \right) \frac{x_0 \left( x_0 -q^2
\right)^2}{\left( x_0^3-2q^2 \right)^3}
\end{eqnarray}
which reduce to the Schwarzschild coefficients when
$q\longrightarrow 0$. From the equation $\alpha=0$, we derive the
radius of the photon sphere
\begin{equation}
x_m=\frac{3}{4} \left(1+\sqrt{1-\frac{32q^2}{9} }\right),
\end{equation}
which yields
\begin{eqnarray}
&\beta_m & =\left[-9+32q^2-144q^4+512q^6+  \sqrt{9-32q^2} \right.
\nonumber
\\%
&& \cdot  \left. \left( 3+16q^2-80q^4 \right) \right] \left[8
\left(q-4q^3 \right)\right]^{-2}.
\end{eqnarray}

The regular term $b_R$ cannot be calculated analytically, but we
can expand the integrand in (\ref{IR}) in powers of $q$ and
evaluate the single coefficients. We get
\begin{equation}
b_R=b_{R,0}+b_{R,2} q^2+O(q^4),
\end{equation}
where $b_{R,0}$ is the value of the coefficient for an uncharged
black hole, calculated in the previous subsection and given by Eq.
(\ref{bR Sch}); the correction is quadratic in the charge $q$ of
the black hole. Its coefficient is
\begin{equation}
b_{R,2}=\frac{8}{9}\left\{ \sqrt{3}-4+\log \left[6 \left(2-
\sqrt{3} \right) \right] \right\}=-1.5939.
\end{equation}

It is very easy to calculate further terms in the expansion of
$b_R$, deriving analytical formulae which prove to be very
accurate even for large values of $q$.

Following Sect. \ref{Sec Strong Field}, we calculate the
coefficients for the formula of the deflection angle
\begin{eqnarray}
& \overline{a} &= \frac{x_m \sqrt{x_m-2q^2}}{\sqrt{\left(3-x_m
\right) x_m^2-
9q^2 x_m +8q^4}}\\%
& \overline{b}&=-\pi +b_R+ \overline{a} \log 2\left( x_m-q^2 \right)^2 \cdot \nonumber \\%
&&  \cdot \frac{ \left[ \left(3-x_m \right) x_m^2-9q^2 x_m +8q^4
\right]
}{\left(x_m-2q^2 \right)^3 \left(x_m^2-x_m+q^2\right) } \\%
& u_m &=\frac{\left(3+\sqrt{9-32q^2} \right)^2}{4\sqrt{2}
\sqrt{3-8q^2+ \sqrt{9-32q^2}}}.
\end{eqnarray}

In \cite{ERT}, the coefficients $a$ and $b$ were calculated
numerically. Here, by our general method, we have been able to
derive the coefficient $\overline{a}$ for the logarithmic
divergence exactly and find a formula for $\overline{b}$ which is
valid up to second order in $q$, indicating the way to extend it
to an arbitrary order.

We notice that the radius of the photon sphere $x_m$ decreases as
the charge increases, but becomes imaginary only for
$q=\frac{3}{4\sqrt{2}}>\frac{1}{2}$; i.e., even when there is no
horizon, a photon can be captured by the gravitational field of a
hypothetical object with a charge greater than the critical value.

\begin{figure}
\resizebox{\hsize}{!}{\includegraphics{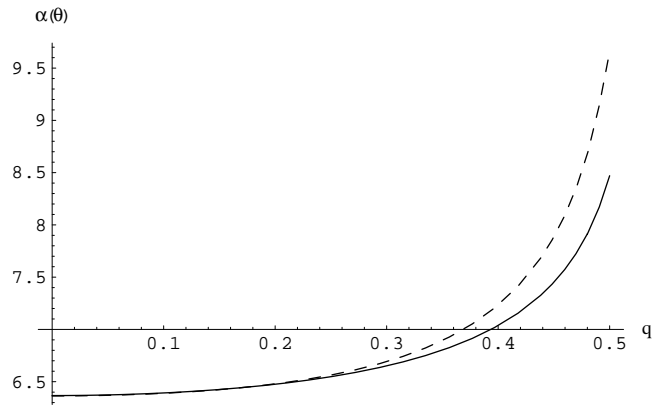}}
 \caption{Deflection angles in Reissner-Nordstrom metric evaluated at
 $u=u_m+0.003$ as functions of $q$. The solid line is the exact deflection angle;
 the dashed line is the strong field limit with $b_R$ truncated to second order in $q$.}
 \label{Fig Rei angle}
\end{figure}

In Fig. \ref{Fig Rei angle}, we evaluate the deflection angle at
$u=u_m+0.003$ as a function of $q$. The plot shows that, just
using the first correction to $b_R$, we get an excellent
approximation to the deflection angle. In fact, we see that up to
$q=0.3$, the error in the position of the outer image, calculated
using the strong field limit, stays below $4\%$.

\begin{figure}
\resizebox{\hsize}{!}{\includegraphics{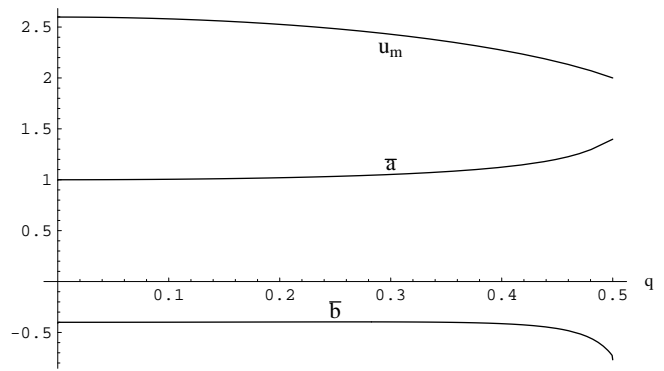}}
 \caption{Coefficients of the strong field limit in Reissner-Nordstrom metric
 as functions of $q$.}
 \label{Fig Rei coefficients}
\end{figure}

Finally, in Fig. \ref{Fig Rei coefficients}, we plot the
coefficients of the strong field limit as functions of $q$,
calculating the full $\overline{b}$ numerically. We see that
$\overline{a}$ and $\overline{b}$ deviate from the corresponding
Schwarzschild values as the charge increases. As we shall see in
Sect. \ref{Sec obs}, the strong field limit coefficients are
directly connected to the observables. It is then possible, in
principle, to distinguish a Reissner-Nordstrom black hole from a
Schwarzschild black hole, using strong field gravitational
lensing.

\subsection{Janis-Newman-Winicour lensing}

The spherically symmetric solution to the Einstein massless scalar
equations ($R_{\mu\nu}=\Phi_{,\mu}\Phi_{,\nu}$,
$\Phi_{;\mu}^{;\mu}$) can be written in Janis-Newman-Winicour
(JNW) coordinates \cite{Brans}
\begin{eqnarray}
&& A(x)=\left(1-\frac{1}{x} \right)^\gamma \\%
&& B(x)=\left(1-\frac{1}{x} \right)^{-\gamma} \\%
&& C(x)=\left(1-\frac{1}{x} \right)^{1-\gamma}x^2 \\%
&&\Phi(x)=\frac{q}{2\sqrt{M^2+q^2}}\log \left(1-\frac{1}{x}
\right) \\%
&&\gamma=\frac{M}{\sqrt{M^2+q^2}}.
\end{eqnarray}
where all distances are measured in terms of $x_s=2\sqrt{M^2+q^2}$
and $q$ is the scalar charge of the central object. This metric
admits a photon sphere external to the static limit when
$\gamma>\frac{1}{2}$, i.e. when $q<M$. We shall thus restrict our
investigations to objects with scalar charge lower than their
mass. In Ref. \cite{VirEll2}, the gravitational lensing of this
object was investigated numerically even when $q>M$. In this
situation, it was shown that a drastically different and
interesting phenomenology shows up.

As in the previous cases, we compute all the coefficients, taking
into account that our metric is not written in standard
coordinates. The coefficients $\alpha$ and $\beta$ are
\begin{eqnarray}
& \alpha &=\left( 2- \frac{2\gamma+1}{x_0} \right)\frac{1}{\gamma
x_0^{\gamma-1} }\left[x_0^\gamma-\left(x_0-1 \right)^\gamma
\right] \\%
& \beta &=-\frac{\left(2\gamma+1\right) \left(\gamma +1 \right)
-2x_0 \left(3\gamma+1 \right) +2x_0^2 }{\left(x_0-1
\right)^\gamma}\cdot \nonumber
\\%
&& \cdot\frac{\left[x_0^\gamma-\left(x_0-1 \right)^\gamma
\right]^2}{2\gamma^2 x_0^\gamma},
\end{eqnarray}
which easily reduce to Eqs. (\ref{alpha sch}) and (\ref{beta sch})
when $\gamma=1$. From the Eq. $\alpha=0$, we derive the radius of
the photon sphere
\begin{equation}
x_m=\frac{2\gamma+1}{2}
\end{equation}
and then
\begin{equation}
\beta_m= \frac{\left[ \left(2\gamma +1\right)^\gamma-\left(2\gamma
-1\right)^\gamma \right]^2}{4\gamma^2 \left( 4\gamma^2 -1
\right)^{2\gamma -1}}
\end{equation}

\begin{figure}
\resizebox{\hsize}{!}{\includegraphics{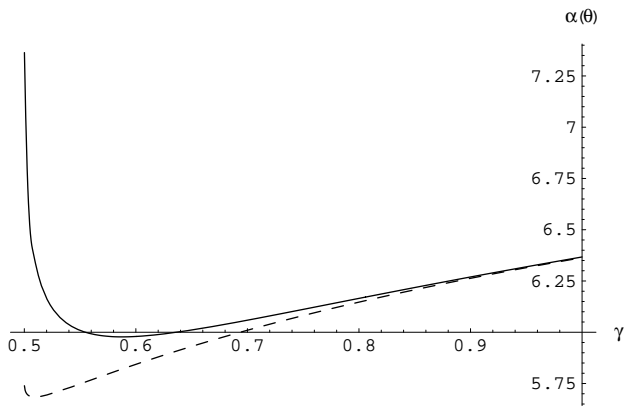}}
 \caption{Deflection angles in a JNW black hole evaluated at
 $u=u_m+0.003$ as functions of $\gamma$. The solid line is the exact deflection angle;
 the dashed line is the strong field limit with $b_R$ truncated to first order in $\gamma$.}
 \label{Fig BD angle}
\end{figure}

\begin{figure}
\resizebox{\hsize}{!}{\includegraphics{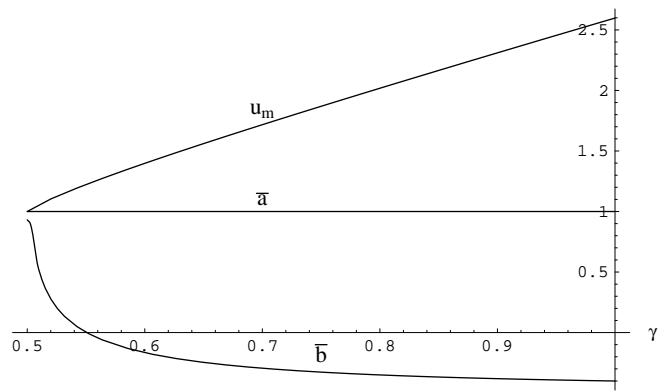}}
 \caption{Coefficients of the strong field limit in a JNW black hole
 as functions of $\gamma$.}
 \label{Fig BD coefficients}
\end{figure}

In the same way as in Reissner-Nordstrom metric, increasing the
charge of the central object, the photons are allowed to get
closer to the centre. When $\gamma=0.5$, $x_m=x_s$ and we should
change our coordinates frame to perform any significant study.

Here as well we cannot solve the integral (\ref{IR}) exactly.
However, we can expand the integrand in powers of $(\gamma -1)$ to
get
\begin{equation}
b_{R}=b_R^0-0.1199 (\gamma-1) +O(\gamma-1)^2
\end{equation}
with $b_R^0$ given by Eq. (\ref{bR Sch}).

Finally we compute the coefficients of the strong field limit
\begin{eqnarray}
& \overline{a} &=1 \\%
& \overline{b} &= -\pi +b_R+ \nonumber \\%
&& +2 \log \frac{\left[ \left(2\gamma
+1\right)^\gamma-\left(2\gamma -1\right)^\gamma \right]^2
\left(2\gamma +1 \right) }{2\gamma^2 \left( 2\gamma -1
\right)^{2\gamma-1}} \\%
& u_m &=\frac{\left( 2\gamma +1
\right)^{\gamma+\frac{1}{2}}}{2\left( 2\gamma -1
\right)^{\gamma-\frac{1}{2}}}
\end{eqnarray}

Surprisingly, the leading order coefficient $\overline{a}$ is the
same as in the Schwarzschild case, in spite of the fact that the
radius of the photon sphere has changed.

Once we fix $u-u_m=0.003$, we see in Fig. \ref{Fig BD angle} that
the deflection angle decreases as we increase the charge (decrease
$\gamma$) until $\gamma$ reaches the value $0.6$, then the
deflection angle increases again. The strong field limit, with
$b_R$ truncated to first order in $\gamma$, at $\gamma=0.7$ is
precise up to $4\%$ in the determination of the outer image.

In Fig. \ref{Fig BD coefficients}, we plot the coefficients of the
strong field limit. $\overline{a}$ is constant, $\overline{b}$
increases as $\gamma$ decreases, but $u_m$ decreases enough to
make the deflection angle decrease at constant $u-u_m$ until
$\gamma$ reaches the value 0.6. Comparing with Fig. \ref{Fig Rei
coefficients}, we can observe that a JNW charge has completely
different effects on the strong field limit coefficients than an
electric charge and can be identified without confusion.

\section{Obervables in the strong field limit} \label{Sec obs}

In Sect. \ref{Sec Strong Field} we have proved that the strong
field limit approximation can be used to obtain a simple and
reliable formula for the deflection angle, which contains a
logarithmic and a constant term. Now we plug the formula (\ref{S F
L theta}) into the lens equation and establish direct relations
between the position and the magnification of the relativistic
images and the deflection angle, calculated according to the
strong field limit.

The lens equation in the strong field limit was derived in Ref.
\cite{BCIS}. It reads
\begin{equation}
\beta=\theta-\frac{D_{LS}}{D_{OS}} \Delta \alpha_n .%
\label{Lens equation}
\end{equation}
where $D_{LS}$ is the distance between the lens and the source,
$D_{OS}=D_{OL}+D_{LS}$, $\beta$ is the angular separation between
the source and the lens, $\theta$ is the angular separation
between the lens and the image, $\Delta \alpha_n =\alpha(\theta)
-2n \pi$ is the offset of the deflection angle, once we subtract
all the loops done by the photon.

To pass from the deflection angle $\alpha(\theta)$ to the offset
$\Delta \alpha_n$, we need to find the values $\theta_n^0$ such
that $\alpha(\theta_n^0)=2n\pi$. Solving Eq. (\ref{alpha theta})
with $\alpha(\theta)=2n \pi$, we find
\begin{eqnarray}
&& \theta_n^0=\frac{u_m}{D_{OL}} \left(1+e_n \right) \label{theta0}\\%
&& e_n=e^{\frac{\overline{b}-2n\pi}{\overline{a}}}.
\end{eqnarray}

The offset $\Delta \alpha_n$ is then found expanding
$\alpha(\theta)$ around $\theta=\theta_n^0$. Letting $\Delta
\theta_n= \theta-\theta_n^0$, we find
\begin{equation}
\Delta \alpha_n=-\frac{\overline{a} D_{OL}} {u_m e_n}
\Delta \theta_n .%
\label{Delta alpha}
\end{equation}

The lens equation becomes
\begin{equation}
\beta=\left( \theta_n^0+\Delta \theta_n \right) + \left( \frac{
\overline{a} D_{OL}}{u_m e_n }\frac{D_{LS}}{D_{OS}} \right) \Delta
\theta_n .%
\label{DTheta lens equation}
\end{equation}

Now we derive the position of all relativistic images, their
magnification and the critical curves of the lens.

The second term in the r.h.s. of Eq. (\ref{DTheta lens equation})
is negligible when compared to the last one (since $u_m\ll
D_{OL}$). Immediately, we find
\begin{equation}
\theta_n = \theta_n^0+\frac{ u_m e_n\left(\beta-\theta_n^0
\right) D_{OS}}{\overline{a} D_{LS}D_{OL}} .%
\label{Images}
\end{equation}
where the correction to $\theta_n^0$ is much smaller than
$\theta_n^0$.

This formula is valid both for the images on the same side of the
source and the images on the opposite side. In fact, to find the
latter, it is sufficient to take a negative $\beta$ in Eq.
(\ref{Images}).

Finally, we have expressed the position of the relativistic images
in terms of the coefficients $\overline{a}$, $\overline{b}$ and
$u_m$. If we manage to determine these coefficients from the
observation of the relativistic images, we are rewarded with
information about the parameters of the black hole stored in them.

The critical curves are just Einstein rings corresponding to a
source perfectly aligned with the lens. Their radius is obtained
putting $\beta=0$ in (\ref{Images}).

Another important source of information is the magnification of
the images which is the inverse of the Jacobian evaluated at the
position of the image. For simplicity, we approximate the position
of the images by $\theta_n^0$, recalling that the correction
provided by Eq. (\ref{Images}) is negligible.
\begin{equation}
\mu_n= \left. \frac{1}{\frac{\beta}{\theta}\frac{\partial
\beta}{\partial \theta}}\right|_{\theta_n^0}.
\end{equation}
We have
\begin{equation}
\left. \frac{\partial \beta}{\partial \theta}\right|_{\theta_n^0}
=1 +\frac{\overline{a} D_{OL}}{u_m e_n}\frac{D_{LS}}{D_{OS}} .
\end{equation}
where the first term is small compared to the second and can be
neglected.

Finally, the magnification is
\begin{equation}
\mu_n=\frac{1}{\left| detJ|_{\theta_n^0}\right|}=
\frac{\theta_n^0}{\beta \left. \frac{\partial \beta}{\partial
\theta}\right|_{\theta_n^0}} =e_n \frac{ u_m^2\left(1+e_n \right)
D_{OS}}{\overline{a} \beta D_{OL}^2 D_{LS}} ,%
\label{Magnification}
\end{equation}
which decreases very quickly in $n$.

Formulae (\ref{theta0}) and (\ref{Magnification}) relate the
position and the magnification to the strong field limit
coefficients. The successive step is to solve the inversion
problem, i.e. we have to find out the most effective way to go
back from measured positions and fluxes to the strong field limit
coefficients, which carry the information about the nature of the
black hole.

The minimum impact parameter can be simply obtained as
\begin{equation}
u_m=D_{OL} \theta_{\infty}
\end{equation}
where $\theta_\infty$ represents the asymptotic position
approached by a set of images, obtained by Eq. (\ref{theta0}) in
the limit $n\rightarrow\infty$.

To obtain the coefficients $\overline{a}$ and $\overline{b}$, we
need to separate at least the outermost image from all the others.
We shall thus consider the simplest situation where only the
outermost image $\theta_1$ is resolved as a single image, while
all the remaining ones are packed together at $\theta_\infty$.

Our observables will thus be
\begin{eqnarray}
&& s=\theta_1-\theta_\infty \\%
&& r=\frac{\mu_1}{\sum\limits_{n=2}^\infty \mu_n},
\end{eqnarray}
which respectively represent the separation between the first
image and the others, and the ratio between the flux of the first
image and the flux coming from all the other images.

The sum of the fluxes of all the set of relativistic images except
the first is
\begin{equation}
\sum\limits_{n=2}^\infty \mu_n=\frac{ u_m^2
D_{OS}e^{\frac{\overline{b}}{\overline{a}}}}{\overline{a} \beta
D_{OL}^2 D_{LS}}
\frac{e^{\frac{4\pi}{\overline{a}}}+e^{\frac{2\pi}{\overline{a}}}+e^{\frac{\overline{b}}{\overline{a}}}}
{e^{\frac{4\pi}{\overline{a}}}-1}.
\end{equation}

We can notice that $e^{\frac{2\pi}{\overline{a}}} \gg 1$ and
$e^{\frac{\overline{b}}{\overline{a}}}$ is of order one, since
$\overline{a}$ and $\overline{b}$ are of order one too. Using
these simple observations, we can simplify our formulae to have

\begin{eqnarray}
&& s= \theta_\infty e^{\frac{\overline{b}}{\overline{a}}
-\frac{2\pi}{\overline{a}}} \\ %
&& r=e^{\frac{2\pi}{\overline{a}}}.
\end{eqnarray}

These two formulae can be easily inverted to give
\begin{eqnarray}
&& \overline{a}=\frac{2\pi}{\log r} \\%
&& \overline{b}=\overline{a} \log \left( \frac{r s}{\theta_\infty}
\right).
\end{eqnarray}
Finally, just measuring an angular separation and a flux ratio, we
are able to reconstruct the full strong field limit expansion of
the deflection angle for the observed gravitational lens.

The coefficients of the strong field limit are constrained by the
characteristics of the metric to be well precise real numbers.
Thus, if strong field gravitational lensing will be detected,
comparing the experimental coefficients with the theoretical
expectations, calculated according to different models, we are
able to identify the nature of the lensing black hole
unambiguously.

\subsection{A numerical example: lensing by the galactic
supermassive black hole}

It is significant to consider a realistic case where we can
discuss the instrumental sensitivity required to detect
relativistic images and possibly distinguish between different
black holes through the reconstruction of the strong field limit
coefficients.

The centre of our galaxy is believed to host a black hole with
mass $M=2.8 \times 10^6 M_{\odot}$ \cite{Richstone}. The lensing
of a background source by this supermassive black hole was
discussed in detail by Virbhadra \& Ellis \cite{VirEll}. Taking
$D_{OL}=8.5$kpc as the distance between the sun and the center of
Galaxy, they found that the separation between each set of
relativistic images with respect to the central lens would be
$\theta_\infty \sim 17$ microarcsecs. In principle, such a
resolution is reachable by actual VLBI projects, but we must be
aware that the disturbances intrinsic in such observations (mainly
due to extinction and emission by accreting matter), would make
the identification of the relativistic images very difficult, as
already pointed out in Ref. \cite{VirEll}.

In Tab. \ref{Table}, following this line, we estimate the
quantities we need for a complete strong field limit
reconstruction in different situations, starting from a simple
Schwarzschild black hole and then going to black holes with
different values of the electric charge and JNW charge.

\begin{table*}
\begin{tabular}{|l|c|c|c|c|c|c|c|c|c|}
  \hline
    & Schwarzschild & \multicolumn{4}{|c|}{Reissner - Nordstrom} & \multicolumn{4}{|c|}{Janis-Newman-Winicour} \\
    \cline{3-10}
    &  & $q=0.1$ & $q=0.2$ & $q=0.3$ & $q=0.4$ & $\gamma=0.9$ & $\gamma=0.8$ & $\gamma=0.7$ & $\gamma=0.6$\\
    \hline
  $\theta_\infty$ ($\mu$arcsecs) & 16.87 & 16.76 & 16.41 & 15.78 & 14.76 & 16.67 & 16.38 & 15.93 & 15.13 \\
  $s$ ($\mu$arcsecs) & 0.0211 & 0.0216 & 0.0234 & 0.0275 & 0.038 & 0.0213 & 0.0216 & 0.0222 & 0.0239 \\
  $r_m$ (magnitudes) & 6.82 & 6.79 & 6.69 & 6.49 & 6.07 & 6.82 & 6.82 & 6.82 & 6.82 \\
  $u_m/R_S$ & 2.6 & 2.58 & 2.53 & 2.43 & 2.27 & 2.57 & 2.52 & 2.45 & 2.33 \\
  $\overline{a}$ & 1 & 1.005 & 1.02 & 1.052 & 1.123 & 1 & 1 & 1 & 1 \\
  $\overline{b}$ & -0.4002 & -0.3993 & -0.3972 & -0.3965 & -0.4136 & -0.3808 & -0.35 & -0.2945 & -0.1659 \\
\hline
\end{tabular}
\caption{ \label{Table} Estimates for the main observables and the
strong field limit coefficients for the black hole at the center
of our galaxy in different hypotheses for the spacetime geometry.
$\theta_\infty$ and $s$ are defined in the text; $r_m$ is $r$
converted to magnitudes: $r_m= 2.5 \; \mathrm{Log} \; r$; $u_m$,
$\overline{a}$ and $\overline{b}$ are the strong field limit
coefficients; $R_S=\frac{2GM}{c^2}$ is the Schwarzschild radius.}
\end{table*}

Looking at Tab. \ref{Table}, we can make some considerations of
different order. Indeed, the easiest parameter to evaluate is the
minimum impact parameter $u_m$, since a microarcsec resolution is
reachable in the next years. This information alone, can already
distinguish between a Schwarzschild or other types of geometry. In
fact, since the total mass and the distance of the black hole are
known to a reasonable accuracy (and possibly will be even better
fixed in the next years), an $u_m$ smaller than predicted would
signal that the structure of spacetime close to the central black
hole is not described by Schwarzschild solution. On the other
hand, if $u_m$ is compatible within experimental uncertainties
with Schwarzschild case, we would set an upper limit for the
parameters describing other black holes, such as an electric or a
scalar charge.

In a second extent, to fit all the strong field limit coefficients
into any black hole model, we need to separate at least the
outermost relativistic image from the others. We see that this can
be done only increasing optical resolution at least by two orders
of magnitude with respect to actual observational projects.
Therefore, with these numbers, it seems that we are forced to wait
for further technological developments. However, given the
evolution rate of astronomical facilities in the last twenty
years, it is not unthinkable that these two orders of magnitude
will be reached within a not so far future.

Black holes are also present in the bulge of other galaxies. As
far as we have investigated, in the best cases, the instrumental
resolutions needed for strong field gravitational lensing are
about the same as for the Milky Way black hole. So, the
determination of $u_m$ by next future observations would possibly
confirm or disprove Schwarzschild geometry for several
extra-galactic black holes too. The complete determination of the
strong field limit coefficients instead remains a long term
project, unless more favourable astrophysical situations emerge.

\section{Summary}

Gravitational lensing is undoubtedly  a potential powerful tool
for the investigation of strong fields. By general arguments we
have shown that the deflection angle diverges logarithmically as
we approach the photon sphere. We have drawn a general method to
compute the coefficient of the leading order divergent term and
the first regular term. When the latter cannot be calculated
analytically, we have seen that it can be well approximated by a
simple series expansion starting from Schwarzschild spacetime.

We have applied our method to Schwarzschild, Reissner-Nordstrom
and Janis-Newman-Winicour black hole, explicitly calculating and
plotting the strong field limit coefficients.

Of course, it is possible to apply the strong field limit, in the
form given in this paper, to any spherically symmetric metric
representing a black hole. In this way, it is possible to compare
the gravitational lensing behaviour of these objects in different
theories of gravitation. In principle, the extension to
non-spherically symmetric and rotating black holes is possible.
However, the dependence of the deflection angle on more than one
variable can put severe obstacles in the way of analytic solutions
of the problem. Nevertheless, this is indeed another important
point which needs to be investigated to complete the picture of
black hole lensing.

Differences in the deflection angle are immediately reflected on
the relativistic images. If the mass and the distance of the lens
is known, then the detection of any set of relativistic images
would immediately check the Schwarzschild geometry. VLBI should be
able to provide an observational answer, if the relativistic
images are not hidden behind environmental noise.

Furthermore, if the outermost image is resolved from the others,
it is then possible to fully reconstruct the strong field limit
coefficients and select a precise black hole model. Our present
observational facilities are not so far from the required
resolutions, which, for the galactic black hole, are of the order
of 0.01 microarcsecs. As a long term project the detection of the
outermost image stands as a very interesting, non-trivial
challenge for future technology.

Strong field limit represents an important step in the
construction of a robust theoretical scheme connecting the
gravitational lensing with the strong field properties. By a
simple and reliable expansion, it clarifies the whole
phenomenology and the differences between various models that
should be expected in the appearance of relativistic images. If
these so elusive features will be detected, we will finally have a
way to effectively discriminate between alternative theories of
gravitation and grow in our knowledge of spacetime.

\bigskip

\begin{acknowledgments}
I would like to thank Gaetano Lambiase and Salvatore Capozziello
for helpful comments and discussions.
\end{acknowledgments}

\end{document}